\documentclass{article}
\usepackage{smc2020}
\usepackage{times}
\usepackage{ifpdf}
\usepackage[english]{babel}
\usepackage{cite}

\newcommand\Mark[1]{\textsuperscript#1}

\def\papertitle{REAL-TIME VISUALISATION OF FUGUE PLAYED BY A STRING QUARTET}
\def\firstauthor{Olivier Lartillot \Mark{1}}
\def\secondauthor{Carlos Cancino-Chac\'on \Mark{2} \Mark{1}}
\def\thirdauthor{Charles Brazier}

\newif\ifpdf
\ifx\pdfoutput\relax
\else
   \ifcase\pdfoutput
      \pdffalse
   \else
      \pdftrue
\fi

\ifpdf 
  \usepackage[pdftex,
    pdftitle={\papertitle},
    pdfauthor={\firstauthor, \secondauthor, \thirdauthor},
    bookmarksnumbered, 
    pdfstartview=XYZ 
   ]{hyperref}

  \usepackage[pdftex]{graphicx}
  \graphicspath{{./figures/}}
  \DeclareGraphicsExtensions{.pdf,.jpeg,.png}

  \usepackage[figure,table]{hypcap}

\else 
  \usepackage[dvips,
    bookmarksnumbered, 
    pdfstartview=XYZ 
  ]{hyperref}  

  \usepackage[dvips]{epsfig,graphicx}
  \graphicspath{{./figures/}}
  \DeclareGraphicsExtensions{.eps}

  \usepackage[figure,table]{hypcap}
\fi

\hypersetup{
    colorlinks,%
    citecolor=black,%
    filecolor=black,%
    linkcolor=black,%
    urlcolor=black
}

\title{\papertitle}

 \threeauthors
   {\firstauthor} {\Mark{1} RITMO Centre for Interdisciplinary Studies\\ in Rhythm, Time and Motion\\
   University of Oslo \\ %
     {\tt \href{mailto:olivier.lartillot@imv.uio.no}{olivier.lartillot@imv.uio.no}}}
   {\secondauthor} {\Mark{2} Austrian Research Institute for \\Artificial Intelligence \\ %
     {\tt \href{mailto:carlos.cancino@ofai.at}{carlos.cancino@ofai.at}}}
   {\thirdauthor} { Institute of Computational Perception \\Johannes Kepler University Linz\\ %
     {\tt \href{mailto:charles.brazier@jku.at}{charles.brazier@jku.at}}}

\begin{document}
\capstartfalse
\maketitle
\capstarttrue
\begin{abstract}
We present a new system for real-time visualisation of music performance, focused for the moment on a fugue played by a string quartet. The basic principle is to offer a visual guide to better understand music using strategies that should be as engaging, accessible and effective as possible. The pitch curves related to the separate voices are drawn on a space whose temporal axis is normalised with respect to metrical positions, and aligned vertically with respect to their thematic and motivic classification. Aspects related to tonality are represented as well. We describe the underlying technologies we have developed and the technical setting. In particular, the rhythmical and structural representation of the piece relies on real-time polyphonic audio-to-score alignment using online dynamic time warping. The visualisation will be presented at a concert of the Danish String Quartet, performing the last piece of \emph{The Art of Fugue} by Johann Sebastian Bach.
\end{abstract}

\section{Introduction}\label{sec:introduction}

In this article, we present a new project of real-time visualisation of music performance. It is part of a broad objective, by the first author and especially within the context of his \emph{MIRAGE} project\footnote{\url{https://www.uio.no/ritmo/english/projects/mirage}}, to design tools to make music more easy to understand and more engaging, especially for non-expert listeners.

The presented project is focused on a String Quartet interpretation of the last piece of \emph{The Art of Fugue} by Johann Sebastian Bach. In order to provide a very rich musicological visualisation of the piece, we decided to prepare the analysis in advance and to synchronise the live performance with the help of score-following technologies.

The paper is organised as follows. Section \ref{previous} gives a brief overview of previous works, including ours. Section \ref{philo} formalises the underlying principles founding our proposed approach. Section \ref{scenario} describes what the visualisation is about, concretely. Section \ref{technical} details the underlying technical aspects and especially the score-following approach.

In the following, a distinction needs to be made between pitch tra\emph{ck}ing (detecting pitch height from audio, frame by frame, as described in Section \ref{tracking}) and pitch tra\emph{c}ing (drawing the pitch line in the visualisation, as described in Section \ref{tracing}).

\section{Previous Works in Music Visualisation} \label{previous}

\subsection{Brief State of the Art}

There is a very large number of approaches that have been proposed for video visualisation of music as it evolves along time. For instance, there has been discussion about the application of music-colour synaesthetic mappings in live ensemble performance \cite{Ng}. A system architecture has been proposed to build real-time music visualisations rapidly, with the view to provide a sense of musical experience to the users \cite{Nanayakkara}.

\subsection{Our Previous Works in Music Visualisation}

This interest in music visualisation stems from the first author's work in audio and music feature extraction and especially the design of \emph{MIRtoolbox} \cite{mirtoolbox}. This toolbox was initially conceived with the initial aim to better understand the affective and emotional impact of music \cite{eerola}. First video visualisations were made in 2012\footnote{\url{https://www.youtube.com/watch?v=H_SpFh2SPmg}}, so far focusing on timbral and dynamical aspects of audio recordings. Another visualisation\footnote{\url{https://www.youtube.com/watch?v=8S7LyEkzWGE}} was aimed at highlighting audio and musical features contributing to emotions dynamically expressed by music \cite{power}. Recently, we designed a new type of visualisation with transformation of video shootings based on audio analysis, and applied to the production of a music 
video\footnote{\url{https://www.youtube.com/watch?v=L1FaxT74hTk}}.

\section{General Philosophy} \label{philo}

This visualisation project is guided by a few general desiderata, presented below, that may look incompatible at first sight, due to conflicting constraints. An additional principle addresses this conundrum.

\subsection{Desiderata} 

\subsubsection{Accessible and Self-Explanatory} \label{accessible}

The visuals should be understandable by a largest public. Hence no music expertise should be required. In particular, music score representations should not be displayed. The visuals should also be self-explanatory. No verbal or text explanation should be necessary for the viewers to grasp the logic.

\subsubsection{Expressive while Thrifty} \label{minimal}

The visuals should be as impactful as possible while at the same time being minimalist with regard to the use of visual strategies. For instance, there is no need to use colors, forms or textures if they are not motivated by a particular need to convey a particular musical characteristic.

\subsubsection{Conveying the Richness of Music} \label{richness}

One aspiration is to  offer a visual equivalent of the way listeners --- and especially experienced music lovers --- perceive, grasp and appreciate music through their ears. Music is generally a rich experience consisting of a flow of events appearing together or successively and building rich interactions. Listeners are immersed by this flow of information, inciting to participate and move with the music. One objective is to convey this experience visually.

\subsection{Addressing Conflicts in the Desiderata}

\subsubsection{A Visual Complexity Assimilable Gradually} \label{gradually}

Clearly, describing the whole musical experience visually may lead to a very complex representation that might be unfathomable at first sight. Although complex stimuli might appeal to the spectators' interest, in order to correctly address desideratum \ref{accessible}, there should be a guiding thread within the flow of stimuli that the spectator can rely on to progressively grasp what is happening and discover little by little the logic of the whole visualisation.

\section{Visualisation scenario} \label{scenario}

The proposed scenario exploits the particular characteristics of string quartet fugues , where each of the four instruments plays a monody and where the subtle changes in pitch, dynamics and timbre play an important role. In a fugue, a \emph{subject} --- a musical theme --- is introduced at the beginning in imitation --- i.e., repeated successively on each voice at different registers --- and reappears frequently in the course of the composition. The particular piece under study, \emph{Contrapunctus XIV}, is actually structured into three sections, each introducing a new subject. 

\subsection{Polyphonic Pitch Tra\emph{c}ing} \label{tracing}

Each instrument plays a single monodic line where successive notes are clearly separated. The slow tempo and the rare use of small rhythmic values such as sixteenth notes also contribute to the clarity of the melodic lines, and to a focus on  pitch, dynamic and timbral shaping of each successive notes. As such, it seemed relevant to simply display pitch curves, resulting directly from the temporal tracking of the fundamental frequency. The pitch curves are progressively drawn from left ro right as time goes by. Each instrument generates its own pitch curve.

\subsubsection{Rhythm Quantisation} \label{quantisation}

As we will see in Section \ref{paradigmatic}, the pitch curves will be superposed so that similar motives could be compared. For that reason, in order to allow their temporal alignments, the horizontal axis directing the pitch curves needs to be temporally quantised, in the sense that it should be graduated in rhythmic values, as indicated in the score. The rhythmic quantisation of the live performance is made possible thanks to score following capabilities, described in section \ref{follower}.

The fact that the tempo is not constant --- and actually fluctuates a lot throughout the piece --- raises a technical issue: How to progressively draw a pitch curve along this quantised axis if there is uncertainty concerning the quantised position of each successive instant?

For each note, a hypothetical tempo --- i.e. a duration in seconds for that note --- is inferred. In one strategy, $S_1$, the tempo is assumed to be constant, and the hypothetical tempo is identified with the tempo related to the previous note.

Another strategy, not considered here, would consist in measuring the tempo variation throughout the piece for different performances of the same musicians and use it as a guide for the tempo change expectation. If the tempo changes are guessed correctly, they would first appear in the visualisation, and it would look as if the performers follow the visuals, and in the other case, the behavior of the visualisation might appear random, which might be undesirable. 

Let $t$ be the performance time and $x$ the time axis in the visualisation. Let's consider a note starting at time $t_i$ and ending at time $t_{i+1}$, corresponding to the metrical positions located in abscissae $x_i$ and $x_{i+1}$ . Let's suppose that the chosen tempo for the visualisation is such that the note is expected to end at time $t'_{i+1}$. Three cases need to be considered:
\begin{itemize}

\item If the expected tempo is exact --- i.e., $t'_{i+1} = t_{i+1}$ ---, the curve is drawn in a simple way, by using the simple mapping $m_t$ defined, for $t \in [t_i, t_{i+1}]$ as
\begin{equation} \label{defaultdrawing}
m_t: \theta \in [t_i, t] \mapsto x_i + \frac{\theta - t_i}{t'_{i+1} - t_i} (x_{i+1} - x_i),
\end{equation}
which means that the part of the curve already drawn does not modify along time afterwards. It remains constant afterward, i.e.,  for $t \geq t_{i+1}$,
\begin{equation} \label{enddrawing}
m'_t: \theta \in [t_i, t_{i+1}] \mapsto x_i + \frac{\theta - t_i}{t_{i+1} - t_i} (x_{i+1} - x_i),
\end{equation}

\item If the actual tempo is slower --- i.e., $t'_{i+1} < t_{i+1}$ ---, the curve is drawn in the same way until reaching the expected end of note $t'_{i+1}$, so equation \ref{defaultdrawing} is valid for $t \in [t_i, t'_{i+1}]$. On the other hand, once point $x_{i+1}$ has been reached, the mapping $m_t$ for $t \in [t'_{i+1}, t_{i+1}]$ becomes
\begin{equation}  \label{correctdrawing}
m''_t: \theta \in [t_i, t] \mapsto x_i + \frac{\theta - t_i}{t - t_i} (x_{i+1} - x_i),
\end{equation}
which means that  curve is compressed to accommodate the rest of the note in the same visual space. Once time $t_{i+1}$ is reached, the mapping stabilises to equation \ref{enddrawing}.

\item If on the contrary the actual tempo is faster --- i.e., $t'_{i+1} > t_{i+1}$ ---, the curve is drawn in the same way until reaching the unexpected end of note $t_{i+1}$, so equation \ref{defaultdrawing} is valid for $t \in [t_i, t_{i+1}[$. But as soon as the note is ended, the mapping is corrected and from that point the mapping switches to equation \ref{enddrawing}. For the sake of clarity following desideratum \ref{accessible} and in order to avoid emphasizing excessively this disruption, an abrupt change in the graphical representation should be avoided, and the transition between mappings $m_t$ and $m'_t$ is shown progressively, in accordance with ``fluid'' design principles.

\end{itemize}


\subsubsection{Pitch Height Ordering} \label{pitch}

The disposition of pitch curves on the Y-axis is less of a problem, as a simple conversion of frequencies on a logarithmic scale gives an adequate representation. Besides, accurately representing the relative location in pitch of each instrument enables to convey the registral differences. Moreover, the rare cases of swapping of register order between voices (the second violin crossing for instance the first violin and playing the highest pitch for a few notes) can be immediately noticed in the visualisation. By associating a different color with each instrument (for instance red for cello, black for alto, blue for the second violin and orange for the first violin), this makes this rare disruption in the natural ordering in pitch more noticeable.

Vibrato, portamento and any other form of pitch deviation related to performance are directly exhibited by the pitch curve, and additional treatment to make them accessible to the viewers do not seem necessary in our case.

The actual set of pitch heights used as implicit reference for the pitch space defines both diatonic and chromatic scales. This would lead in theory to a graduation of the pitch axis through a large set of horizontal lines. Following desiderata \ref{accessible}, the display of the whole diatonic or chromatic scale is avoided. Only pitches that are actualized are indicated with horizontal lines. The tonal structuration of this set of lines is further discussed in section \ref{tonal}.

\subsubsection{Dynamics and Timbre}

Amplitude for each pitch curve is represented by controlling the width of the curve. Timbral aspects will also be depicted in future works.


\subsection{Dynamic Paradigmatic Analysis} \label{paradigmatic}

So far the position of the pitch curves on the time axis (as discussed in section \ref{quantisation}) was expressed with respect to the metrical grid given by the notated music. It was implicitly assumed that the X-axis in the visualisation represented the overall temporal evolution of the piece of music from beginning to end. 

Yet such strategy would only show how the music sounds like at each individual instant, without any structured reference with what has been played, apart from the mere display of the left part of the curve before the given time. Music listening is a lot about inferring associations from what we are currently hearing and we previously heard. This leads to the inference of a structural understanding of the piece, and in particular of a thematic and motivic lexicon, where music sequences of various sizes are considered as repetitions of themes and motives. Following desideratum \ref{richness}, this structural information should be depicted in the visualisation.

\begin{figure*}[t]
\centering
\includegraphics[width=1.5\columnwidth]{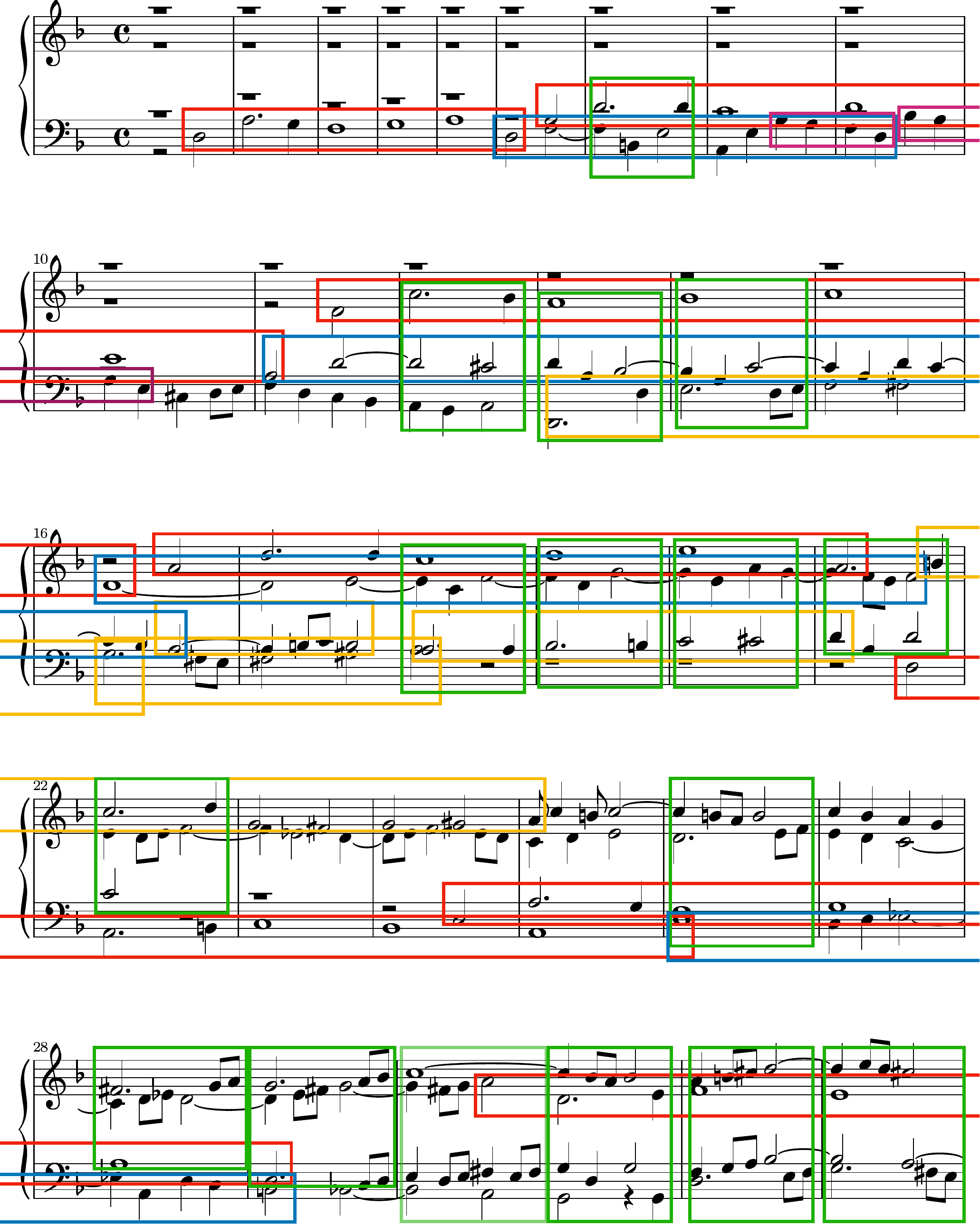}
\caption{Motivic analysis of the beginning of \emph{Contrapunctus XIV}. The first subject is shown in red and its corresponding countersubject in blue. \label{paragfig}}
\end{figure*}

One solution commonly used in music analysis is to add annotations on top of the score --- or in our case, the pitch curve ---, such as boxes around sequences of notes that are repeated. Boxes would be distinguished with respect to color, width, line style, etc., in order to denotate particular theme or motive.
Figure \ref{paragfig} gives an idea of the type of motivic structure that can be found in the beginning of \emph{Contrapunctus XIV}.

The problem with that formalisation is that it makes the visualisation more complicated, more technical, and would demand a lot of additional attention from the spectators, which would contradict desideratum \ref{minimal}. In particular, they would need to find by themselves the position in the score (or the overall set of pitch lines) of each type of boxes, which would quickly become tedious. Besides, it does not seem feasible to display the whole pitch curve on one single screen.

\begin{figure}[t]
\centering
\includegraphics[width=\columnwidth]{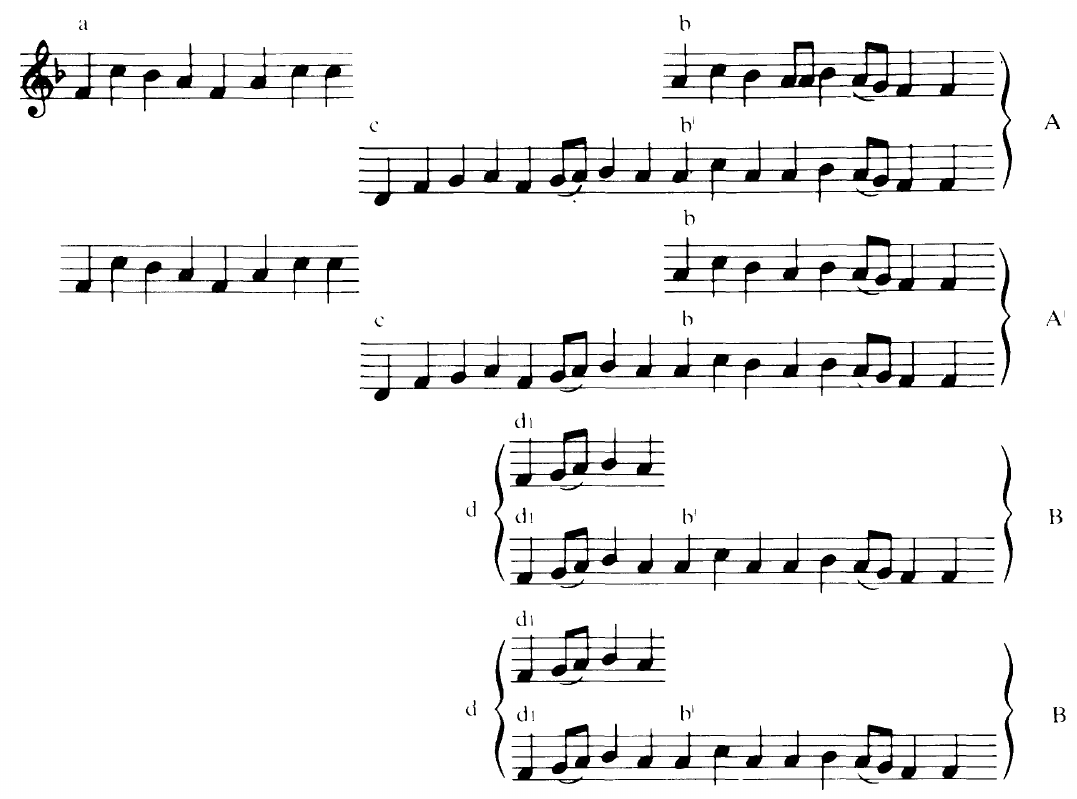}
\caption{Paradigmatic analysis of a \emph{Geisslerlied} \cite{Ruwet}.\label{geisslerlied}}
\end{figure}

A fruitful solution to this problem can be found in the notion of paradigmatic analysis. It was initially a method designed by Levi-Strauss in order to structure texts, and it has then been introduced in music analysis by Nicolas Ruwet \cite{Ruwet} (cf. an example in Figure \ref{geisslerlied}), and used extensively in music semiotics \cite{Nattiez}. The idea is simply to align vertically musical sequences that share a same motivic or thematic identity. By compiling the successive occurrences one below each other, the whole score can be read from left to right and from top to bottom.

We partly follow this strategy by cutting the pitch lines into segments and aligning those corresponding to repetitions of a same motif or theme vertically. Contrary to the paradigmatic analysis, however, we do not use the vertical axis as a way to distribute the sequences over time, because the Y-axis is already used for representing the whole 4-voice polyphony.

Since our visualisation is dynamic, evolving over time, we do not need to care about the readability of the whole representation once everything is drawn, but only about the understandability of the dynamic progression of the visualisation. As such, it is not a problem to superpose the repetitions of a same pattern one on top of each other. If they are transposed (which is generally the case in fugues), the corresponding vertical translation will be clearly shown. Besides, the music operation of inversion (very typical to fugues as well) will lead to the vertical inversion of particular occurrences as well.

Since the pitch curves are not  drawn simply from left to right anymore but may appear at any place in the screen, the part of the curve being extended at a given time should be clearly visible. For that purpose, we use the parameter of line width to also control the recency of each part of the pitch curves. The part of the curve most recently appears thicker while the older sections become thinner and thinner.
Besides, the rightmost point of the pitch curve corresponding to the current instant is highlighted with a clearly visible pointer, which has the color related to the instrument, as mentioned in section \ref{pitch}. 

\begin{figure*}[t]
\centering
\includegraphics[width=\columnwidth]{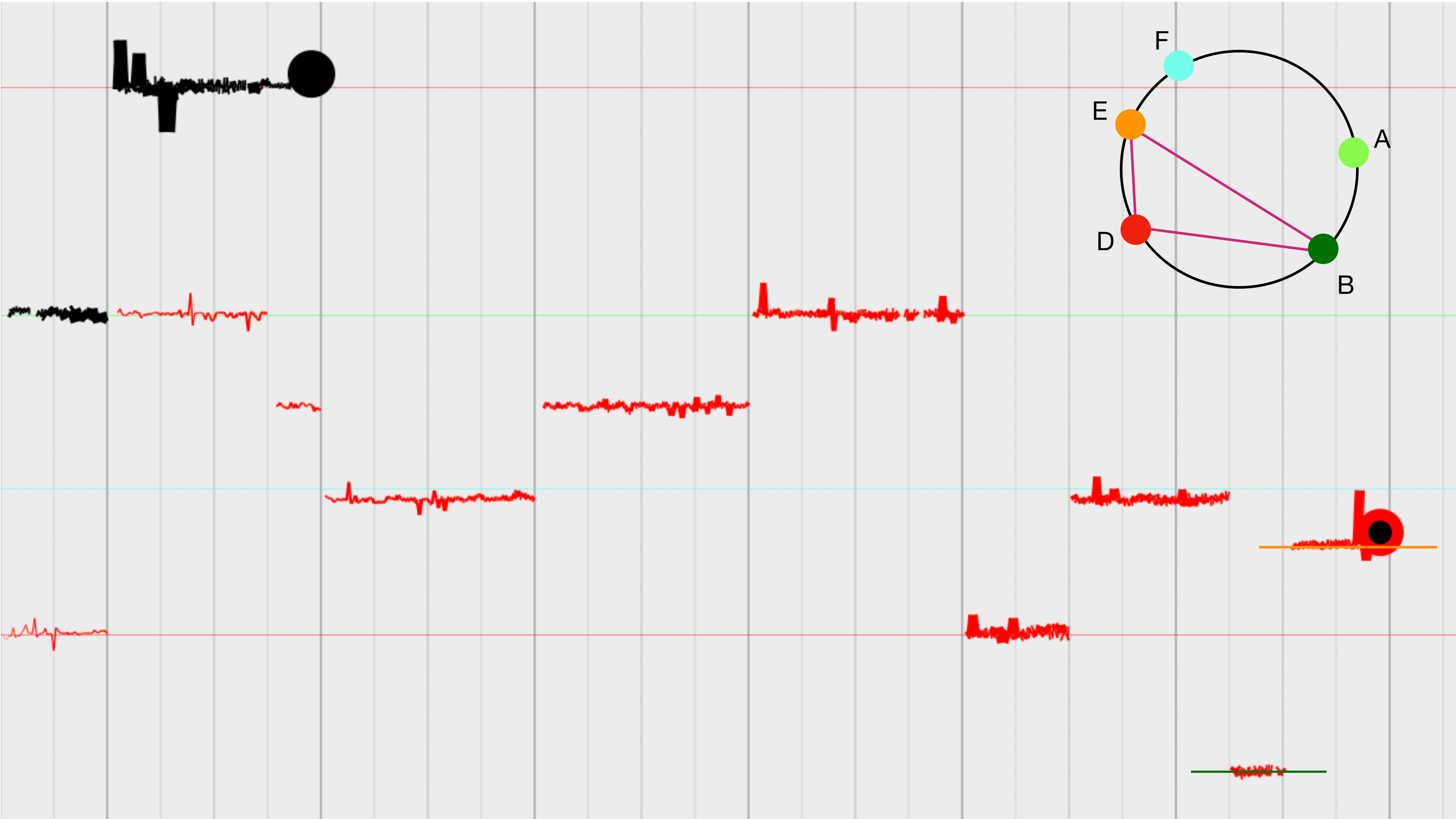}
\includegraphics[width=\columnwidth]{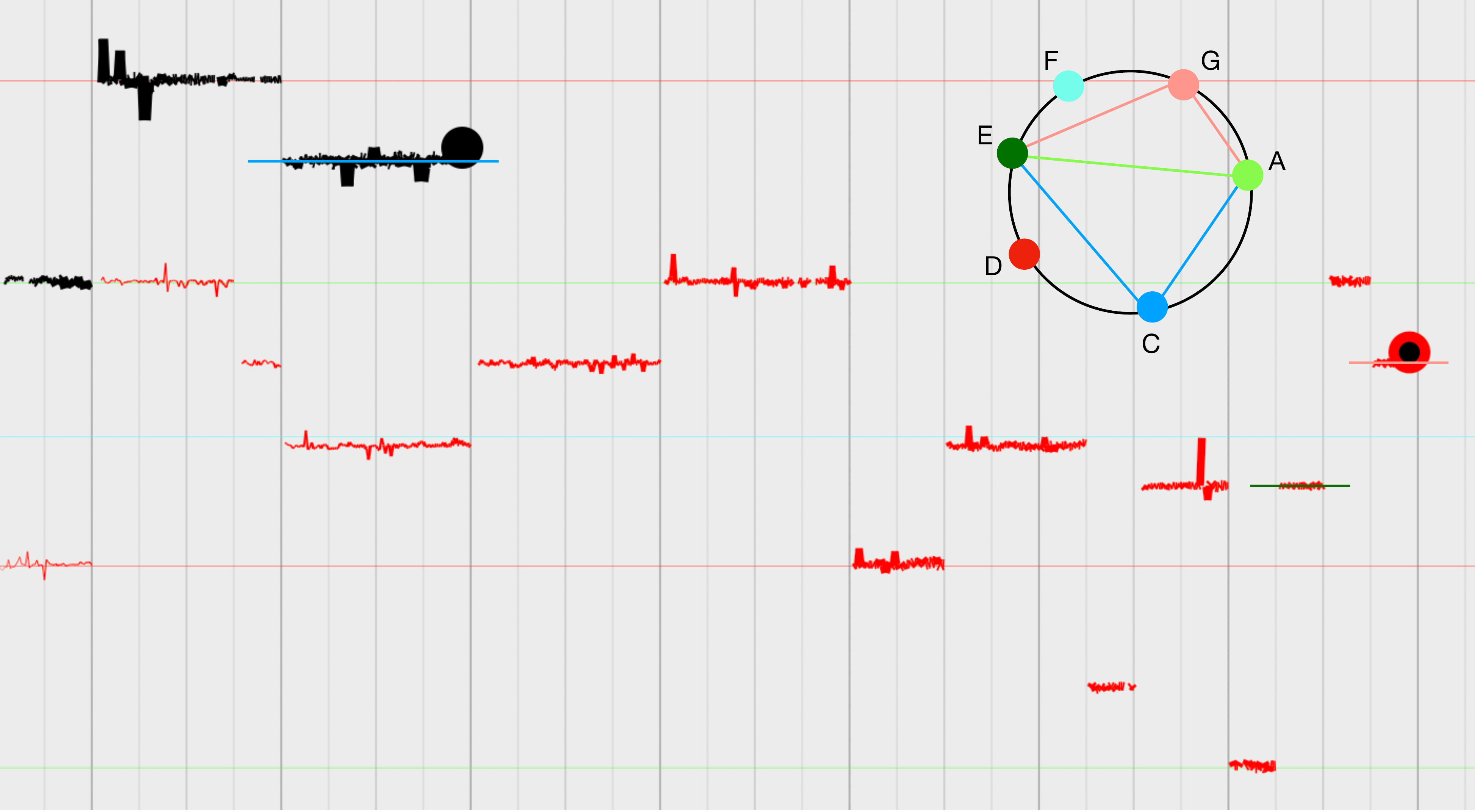}
\includegraphics[width=\columnwidth]{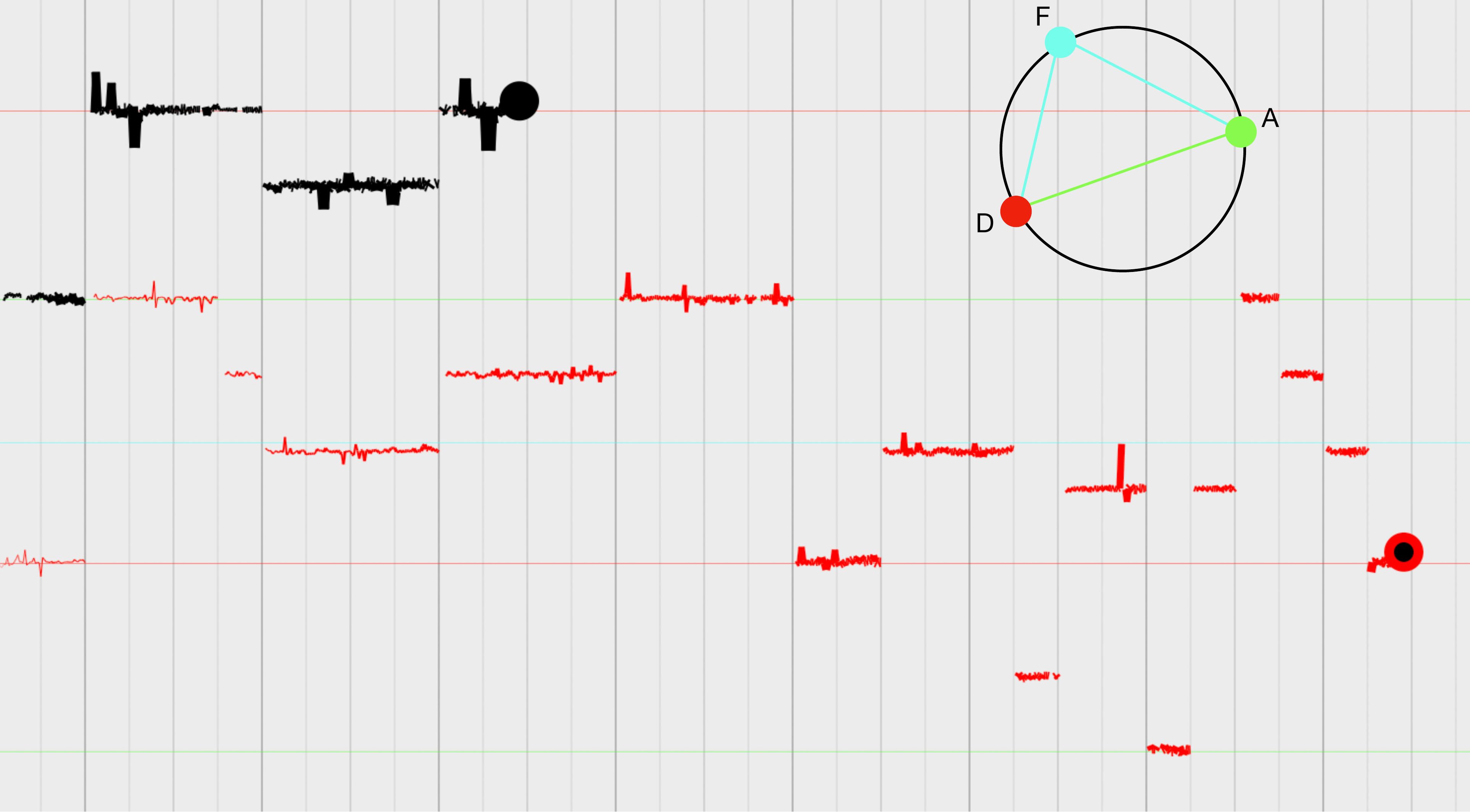}
\includegraphics[width=\columnwidth]{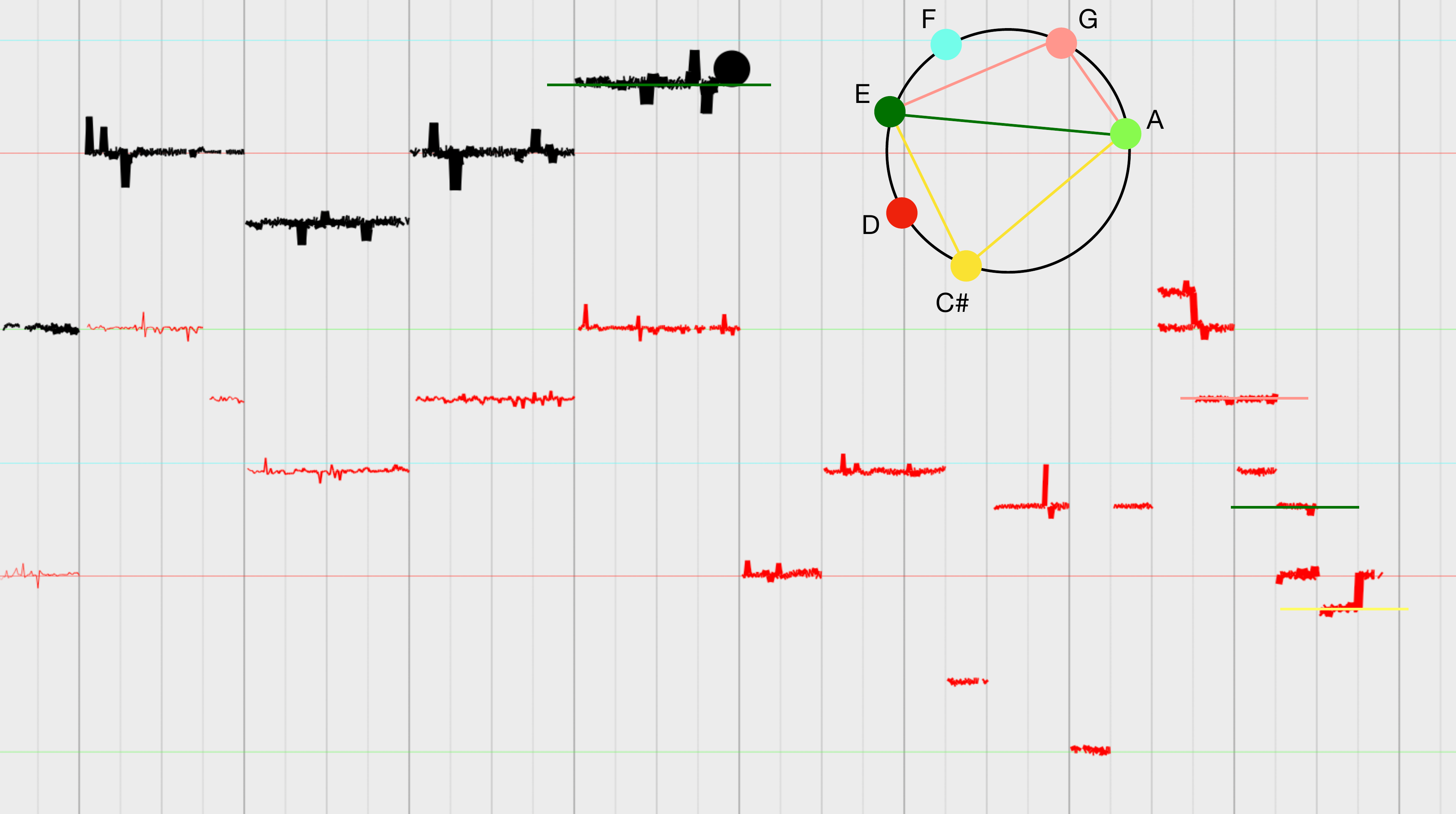}
\includegraphics[width=\columnwidth]{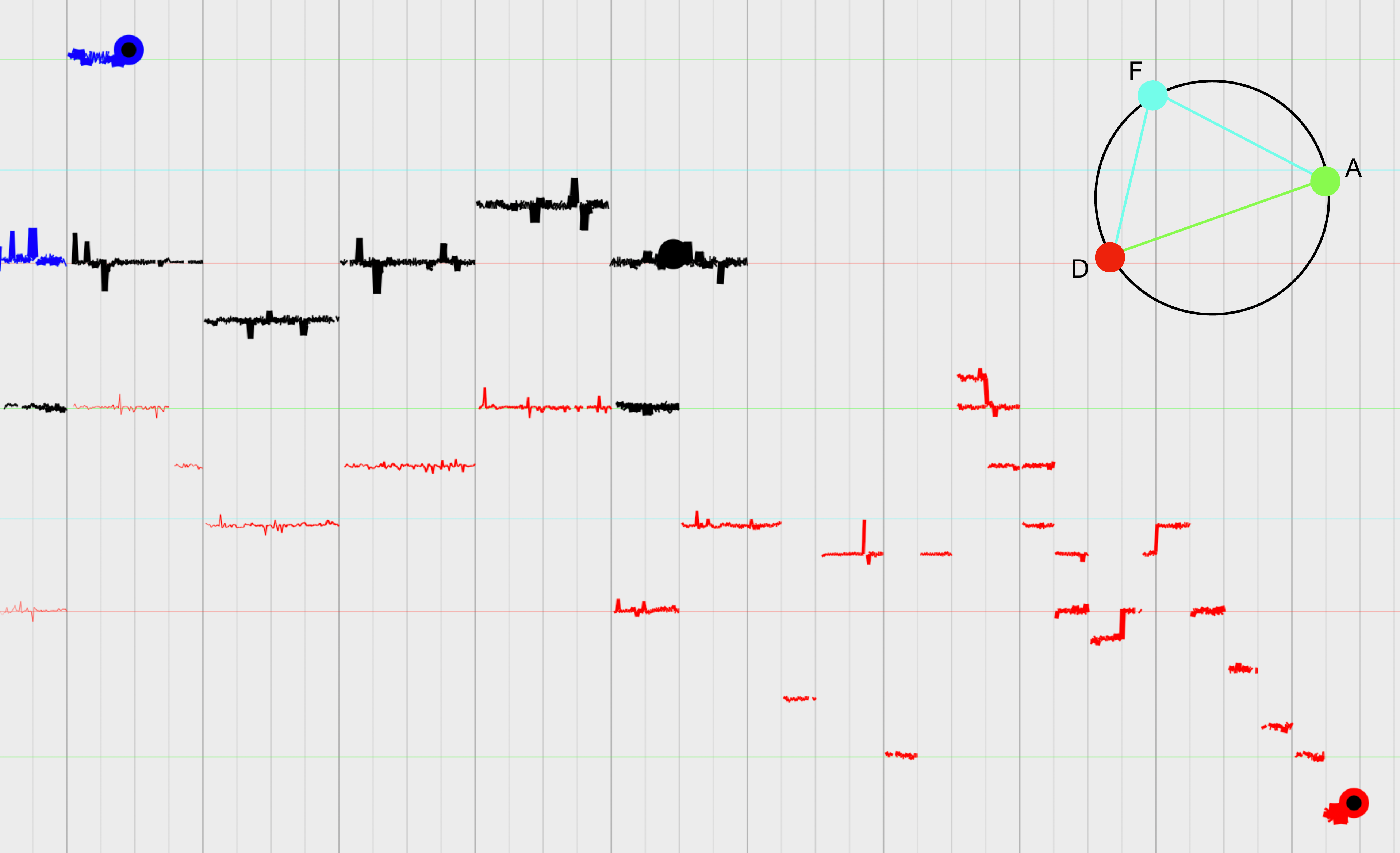}
\includegraphics[width=\columnwidth]{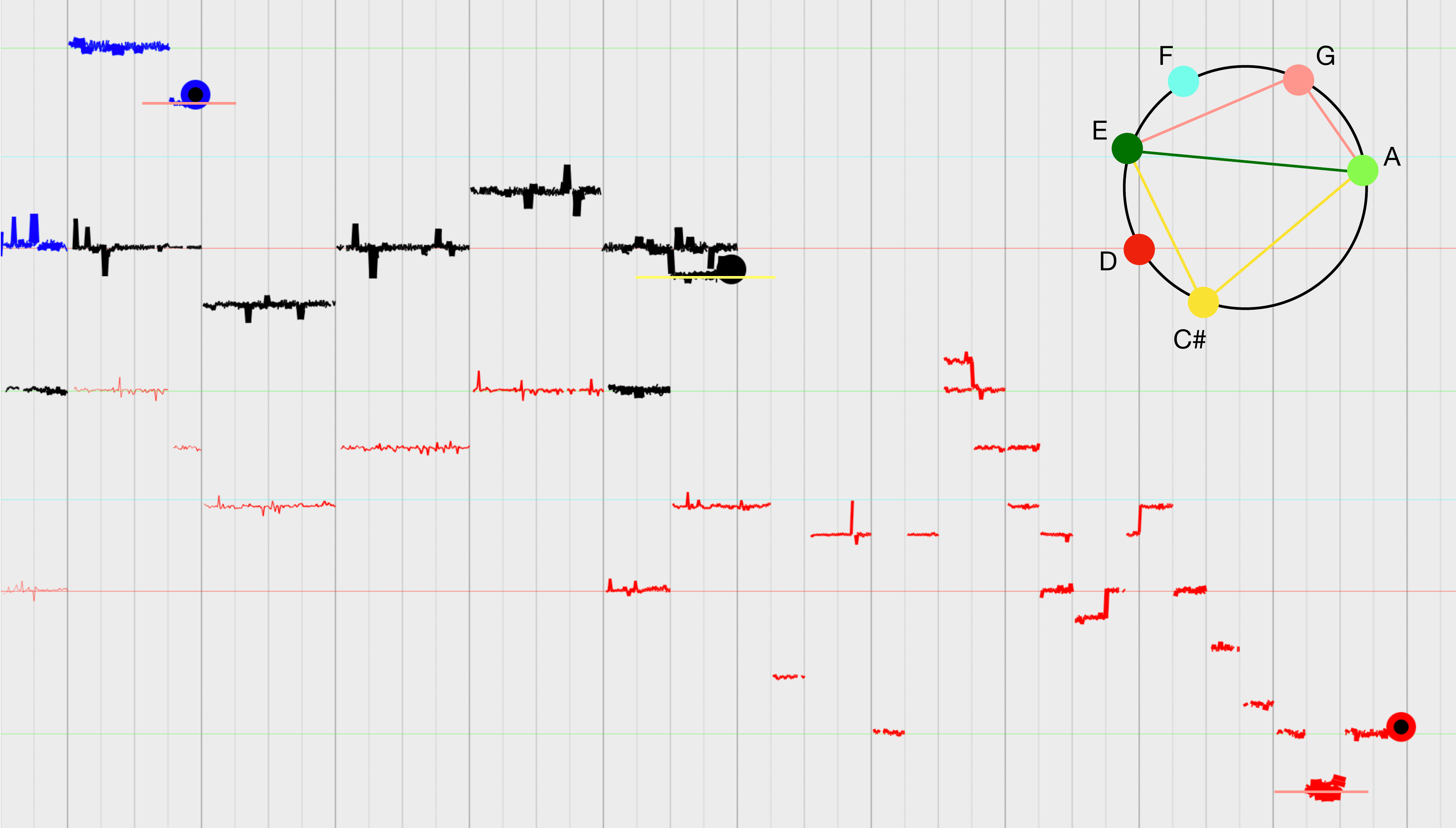}
\caption{Screenshots of the current version of the visualisation showing various moments at the beginning of \emph{Contrapunctus XIV}.
On the top right of each image is added the circular tonal representation, with the main triad D-F-A.
Successively from top to down, left to right: end of bars 7, 8, middle of bar 9, end of bar 10 and middle and end of bar 12.
\label{screenshots}}
\end{figure*}

Figure \ref{screenshots} presents the progressive construction of the para\-digmatic visualisation of the beginning of \emph{Contrapunctus XIV}. The first 5 bars correspond to the first subject, while the rest of the excerpt corresponds to its counter-subject. Only the first occurrence of the counter-subject is shown here. We notice that the successive 1-bar repetition in the bass line in bars 8---9 and 9---10 (purple box in figure \ref{paragfig})  is folded, the second occurrence on top of the first one. Similarly, for the tenor part, bar 12 is an extra bar extending further bar 11, and is therefore folded in the visualisation.

\subsection{Tonal Analysis} \label{tonal}

Concerning the tonal dimension of the music, it is necessary to highlight hierarchies between pitches. As the key of the piece is D minor, it seems reasonable to highlight the main triad D-F-A, which is shown in the visualisation in the form of a horizontal grid. All the lines corresponding to a same pitch class are displayed with a same color (for instance\footnote{Examples of colors are indicated in this paper. They will be tweaked so as to optimise the clarity of the representation and the hierarchy of importance.}, red for D, cyan for F and green for A). Only lines corresponding to pitches that have been actually played are shown, so they  progressively ``light up'' as the music goes.

Attentive spectators will notice that while many entries of the fugue's first subject  go from D (red) to A (green) and back to D, other entries go from A to D and back to A, which has a slight shorter ambitus (fourth instead of fifth, shown respectively by the red and black lines in Figure \ref{screenshots}.). The visualisation thus explicitly reveals the dichotomy between \emph{subject} and \emph{answer} in a fugue.

To make explicit the concept of pitch classes, which plays an important role in tonal analysis, as we will discuss in the next paragraphs, a circular representation of pitch classes is shown in addition to the pitch curve representation. This is illustrated in Figure \ref{screenshots}\footnote{The tonal visualization has not been implemented yet, and is therefore manually superimposed onto the screenshots.}. When the pitch classes D, A and F successively appear at the very beginning of the fugue, they appear at the same time in the circle, using the same color code as previously defined. Here also, this core triad defining the main key of the piece will remain displayed during the whole piece.

One core characteristic of tonality is that the music can be considered as a succession of triads (simple triads or seventh chords). These successive triads are shown as triangles on the circle. Only the current triad is shown. Seventh chords are shown with a triangle representing the main triad, plus a second little triangle between the fifth, seventh and root, as shown in Figure \ref{screenshots}, panels 1, 2 and 4.

For any given triad, if the root does not belong to the D-F-A triad, it is indicated in orange both in the circle and as a horizontal line. Same for the fifth but with a distinction: it is shown in dark green if it is a perfect fifth, and grey in the other cases (not illustrated in the figure). The colors is also used for the side of the triangle linking the root and the fifth. Similar principles are used for the third, but with a distinction between major and minor thirds. Similar principles are defined for the seventh degree.

As soon as the triad has been replaced by a subsequent one, the previous triad is not displayed anymore, neither in the circle, nor as horizontal lines.

The last concept of importance that is visualised is the descending fifth transition, so important in tonal music for its role in cadences but more generally in the ``circle of progression''. Descending-fifth transitions are shown through an actual rotation of the triangle in the circle. At the same time, pitch classes that remain constant through the transition are highlighted as such, while voice leading movements are shown through translations, leading tone going one degree up and the seventh one degree down. Finally, when the descending fifth corresponds to a cadence, arriving to an important tonic chord, the rotation ends with a slightly flamboyant animation.

These tonal considerations will surely appear very arcane at first sight for a non-expert spectator. Following principle \ref{gradually}, these more complex consideration should be displayed in a rather unobtrusive way, available to curious viewers who already understand the rest of the visualisation.

\subsection{General Structure of the Piece}

\emph{Contrapunctus XIV}, is a fugue with three different subjects, defining three successive parts, each starting when the corresponding subject first appears. It should be noted that when the second, or third, subject appears, the previous one(s) can still occur from time to time.

At the beginning, the cello ($c$) plays the first subject $A$ alone, which is displayed on the leftmost part of the screen. Then the alto ($a$) plays the second entry ($A_2$) of $A$ while $c$  plays the countersubject of $A$. It should be noted that in this fugue the countersubjects are not exact repetitions of each other, but can be rather characterised as looser passages structured by specific motivic cells. 
Hence these countersubjects are not represented as single curves but as a succession of segments related to the different motivic cells. For instance in this first countersubject, related to $A_1$, there is a successive repetition of a small pattern with transposition, which is shown on the right of panel 4 of Figure \ref{screenshots}.

The first section continues similarly with the successive entries $A_i$ of $A$ while the other voices continue their entries and develop new motivic materials. All entries of subjects remain visible (while still being progressively dimmed as indicated in section \ref{paradigmatic}). On the contrary,  to simplify the visualisation, some of the entries' continuations and developments completely fade away after some point if they are not deeply related to parts that will appear later in the piece.

The first entry by the second violon ($v_2$) of the second subject $B$ appears, at measure 114, as one of the many developments within the dense network related to subject A. But as $v_2$ continues its long entry solo, it progressively occupies the middle part of the screen, and all the material related to $A$ fades away, so that this new $B$ entry suddenly becomes alone.

It should be noted that $B$ is far longer than the other two subjects, but includes internal motivic repetitions, which are represented as such, through vertical superposition of small lines.

Then the first violin ($v_1$) plays $B_2$ while $v_2$ plays the countersubject of $B_1$, and so on. Then at measure 149, $c$ plays a new entry of $A$ (leading to a progressive visual reactivation of the network of lines related to $A$) while $v_1$ plays
an entry of $B$  and so on.

There are also rare cases of multiple-voice patterns, where several voices play a pattern that is repeated several times successively. This is a rare case where several pitch curves are superposed not because of their thematic identity but because of their simultaneity.

At measure 193, the third subject $C$ is introduced by $a$. This theme, displayed on the right part of the screen, is known as the ``BACH'' theme as it starts with B$\flat$, A, C and B$\natural$. So those letters are displayed next to the pitch curve (as well as on the pitch class circle), which also gives some dramatic effect to the visual.

In measure 233, $v_2$ plays a final entry of $B$ while $c$ starts a final entry of $A$. These two final entries are particularly highlighted in the visualisation. On measure 239,  $v_2$ and $c$ end abruptly, while $a$ continues a short imitation of the ending of $B$ by $v_2$ for just one more measure, and everything stops abruptly, as the composition of the piece is unfinished.

The whole visualisation then fades away with only the four letters B A C H remaining.

\section{Technical Solution} \label{technical}

\begin{figure*}[t]
\centering
\includegraphics[width=1.7\columnwidth]{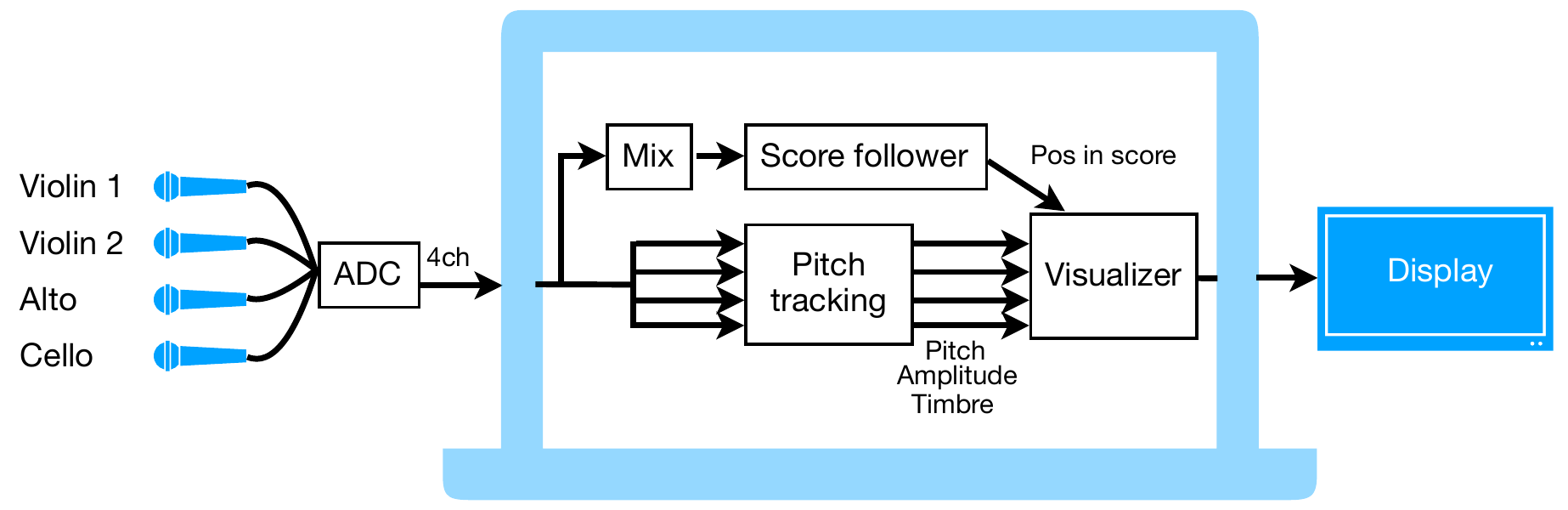}
\caption{Overall architecture of the proposed solution.\label{architecture}}
\end{figure*}

The data-flow diagram  of the proposed solution developed for this project is shown in Figure \ref{architecture}.

Each of the four instruments is captured using one of four DPA 4060 close-up microphones. The audio is digitized and mixed into a single 4-channel audio stream using a Behringer UMC404HD audio interface (to be confirmed), which is plugged via USB to a MacBook Pro.

The software part is integrated into a single Mac app developed in Swift. Each of the four channel is fed to a Pitch Tracking module, which sends information about pitch, amplitude and timbre for each successive frame in each instrument to the Visualisation module. In parallel, the four-channel audio stream is also summed into one mono channel and fed to the Score Following module, which sends information about the current position in the score to the Visualisation module as well.

\subsection{Pitch Tra\emph{ck}ing} \label{tracking}

The audio signal is decomposed into successive frames of length 50 ms and of hop 25 ms. For each successive audio frame, pitch has been so far estimated by computing the autocorrelation function. To speed up the computation, autocorrelation function is computed using Fast Fourier Transform (FFT), using the vDSP library of Apple's \emph{Accelerate} framework. The frequency corresponding to the global maximum of the autocorrelation function is selected as the fundamental frequency.

In the first prototype, the analysis is carried out on synthesized rendering of MIDI files. Future works will focus on the particular characteristics of string instruments. Mechanisms will also be developed to avoid incorrect pitch detection due to leakage between instruments.

Other signal processing processes are under consideration to extract timbral descriptions of the sound.

\subsection{Score Following} \label{follower}

Score following refers to the task and technologies for aligning performances of musical pieces to their score in real-time~\cite{Mueller:2015,Arzt:2016}.
The \emph{dynamic time warping} (DTW) algorithm is the method of choice for performing audio-to-audio alignments \cite{Gadermaier:2019}.
DTW is a general dynamic programming technique for finding the best alignment between two sequences of observations.
For the system presented in this paper, we use the \emph{on-line time warping} (OLTW) algorithm~\cite{Dixon:2005,Arzt:2016}. 
This algorithm adapts DTW to be used in real-time scenarios by computing the alignment incrementally.

We perform audio-to-score alignment by (manually) annotating a live recording of \emph{Contrapunctus XIV} performed by DSQ, which we will refer to in the following as the \emph{reference}.
For this recording, the positions of the beats were annotated using Sonic Visualiser\footnote{\url{https://sonicvisualiser.org}} by two annotators with 10+ years of formal musical education. 
These beat annotations are then averaged across annotators.
As spectral features to compute the input and reference sequences we used  the modified MFCC features described in \cite{Gadermaier:2019}.
These features have been successful in aligning opera in real time~\cite{Brazier:submitted}.

The score follower works as follows:
\begin{enumerate}
\item \textit{Before the beginning of the performance}: Extract spectral features for the annotated reference recording and set the current position of the score follower to 0.
\item \textit{During the performance}:
\begin{enumerate}
\item Mix the the individual tracks into a single (mono) track.
\item For each new audio frame from the mixed input stream, compute the spectral features.
\item Update the current position of the score follower using OLTW.
\end{enumerate}
\end{enumerate}

\subsection{Visualizer}

The module draws the pitch curves --- and their transformations based on amplitude and timbre --- related to each instrument into successive segments, and places them in the screen to form the paradigmatic analysis, as discussed in section \ref{paradigmatic}. Under the pitch curves are also displayed horizontal lines corresponding to important pitch levels related to the tonal analysis, as explained in section \ref{tonal}. The module also draws the tonal circle and the successions of triads and seventh chords. It also displays some particular events such as the pitches B A C H and the particular emphasis on the last entry of subjects $A$ and $B$.

Before the real-time process, the paradigmatic analysis is loaded into a particular data structure, containing for each voice (i.e., each instrument) the list of successive segments. 

Each segment is represented by a triplet $(t_1, t_2, x_1)$, where:
\begin{itemize}
    \item $t_1$ and $t_2$ are the time positions of the start and end of the segment, expressed as positions in the score, and
    \item $x_1$ is the abscissa of the start of the segment in the visual representation.
\end{itemize}{}

The tonal analysis is loaded into another data structure, containing the list of successive chords, with also indications of the falls of fifths. Another list indicates when each horizontal lines corresponding to a pitch from the main triad D, F and A lights up.

\subsection{Control Interface}

A user's interface is also being designed to allow external control of the system. This will enable to control a number of parameters before and during the concert, related in particular to the amplitude of the input signals or the tuning. This will also allow to manually assign the position in the score in case of problems with the score following or of unplanned performance issues from the musicians' side.

\section{Future Works}

 The system is currently under development. 
The visualisation will be premiered at a Danish String Quartet concert, which had to be been postponed to May 22, 2021, at the Concert Hall of the Royal Danish Academy of Music in Frederiksberg. This leaves a comfortable amount of time to improve the visualisation, in particular related to pitch tracking, and  to improve clarity, richness and aesthetic appeal.

The audience will be invited to fill out a questionnaire, to evaluate the fulfillment of desideratum \ref{accessible}, i.e., that the visualisation should be accessible and self-explanatory to a largest public. The visualisation will also be published online, with a corresponding online questionnaire.

In future works, we plan to generalise the system to broader musical contexts, through the integration of polyphony, of various musical forms. We plan also to integrate a polyphonic pitch tracker that would not require the use of multi-channel instrument pickups, but simply a single mono or stereo pickup. The score following might be used as a guiding point for the pitch tracking.

The musicological analysis, here the paradigmatic and tonal analyses, are for the moment performed manually. In longer term we plan to integrate our systems of automated motivic and structural analysis. By integrating also automated transcription systems, the whole process could be carried out without the need of a score and therefore of a score follower.

\begin{acknowledgments}
This work was partially supported by the Research Council of Norway through its Centers of Excellence scheme, project number 262762 and the MIRAGE project, grant number 287152; the European Research Council (ERC) under the European Union’s Horizon 2020 research and innovation programme under grant agreement No. 670035 (project ``Con Espressione"); and the Marie Sk\l{}odowsa-Curie grant agreement number 765068, MIP-Frontiers.

This visualisation project is part of a larger project, the MusicLab5 Copenhagen 2020 event, under the initiative of Simon H{\o}ffding, which has been postponed to 2021. We would like to warmly thank the musicians of the Danish String Quartet and Simon H{\o}ffding for offering us this fantastic opportunity to carry out this project.

The score in Figure \ref{paragfig} was initially typset by Tim Knigge, and made available through the Mutopia project\footnote{\url{http://www.mutopia.org}.}.

\end{acknowledgments} 

\bibliography{smc2020bib}

\end{document}